\documentclass[iop,apj]{emulateapj}
\usepackage{amsmath, amsthm, amssymb,amsfonts}
\usepackage[english]{babel}
\usepackage{epstopdf}
\usepackage{color}
\usepackage{graphicx}
\usepackage[hang,flushmargin]{footmisc}

\usepackage{hyperref}
\definecolor{darkblue}{rgb}{0.0,0.0,0.3}
\hypersetup{colorlinks,breaklinks,
            linkcolor=darkblue,urlcolor=darkblue,
            anchorcolor=darkblue,citecolor=darkblue} 

\newcommand{\pD}[2]{\frac{\partial #2}{\partial #1}}

\newcommand{\D}[2]{\frac{{\rm d} #2}{{\rm d} #1}}

\newcommand\bb[1]{\mbox{\boldmath{$#1$}}}
\newcommand\grad{\bb{\nabla}}

\newcommand\btimes{\,\bb{\times}\,}

\newcommand{\msb}[1]{\bb{\mathsf{#1}}}

\newcommand{\ez}{\hat{\bb{z}}}

\newcommand{\pegpp}{\textsc{Pegasus}\texttt{++}~}
\newcommand{\rmd}{{\rm d}}
\newcommand{\const}{{\rm const}}

\begin{document}

\title{Adaptive critical balance and firehose instability \\ in an expanding, turbulent, collisionless plasma}
\author{A.~F.~A.~Bott$^1$}
\author{L.~Arzamasskiy$^{1,2}$}
\author{M.~W.~Kunz$^{1,3}$}
\author{E.~Quataert$^{1}$}
\author{J.~Squire$^{4}$}
\affiliation{$^1$Department of Astrophysical Sciences, University of Princeton, 4 Ivy Ln, Princeton, NJ 08544, USA}
\affiliation{$^2$Institute for Advanced Study, 1 Einstein Drive, Princeton, NJ 08540, USA}
\affiliation{$^3$Princeton Plasma Physics Laboratory, PO~Box 451, Princeton, NJ 08543, USA}
\affiliation{$^4$Department of Physics, University of Otago, 730 Cumberland Street, Dunedin 9016, New Zealand}
\email{E-mail of corresponding author: abott@princeton.edu}

\begin{abstract}
Using hybrid-kinetic particle-in-cell simulation, we study the evolution of an expanding, collisionless, magnetized plasma in which strong Alfvénic turbulence is persistently driven. Temperature anisotropy generated adiabatically by the plasma expansion (and consequent decrease in the mean magnetic-field strength) gradually reduces the effective elasticity of the field lines, causing reductions in the linear frequency and residual energy of the Alfv\'{e}nic fluctuations. In response, these fluctuations modify their interactions and spatial anisotropy to maintain a scale-by-scale ``critical balance'' between their characteristic linear and nonlinear frequencies. Eventually the plasma becomes unstable to kinetic firehose instabilities, which excite rapidly growing magnetic fluctuations at ion-Larmor scales. The consequent pitch-angle scattering of particles maintains the temperature anisotropy near marginal stability, even as the turbulent plasma continues to expand. The resulting evolution of parallel and perpendicular temperatures does not satisfy double-adiabatic conservation laws, but is described accurately by a simple model that includes anomalous scattering. Our results have implications for understanding the complex interplay between macro- and micro-scale physics in various hot, dilute, astrophysical plasmas, and offer predictions concerning power spectra, residual energy, ion-Larmor-scale spectral breaks, and non-Maxwellian features in ion distribution functions that may be tested by measurements taken in high-beta regions of the solar wind.
\end{abstract}


\maketitle

\section{Introduction} \label{sec:intro}

Many space and astrophysical plasmas are magnetized and weakly collisional, with the Larmor radii of the constituent particles being many orders of magnitude below their Coulomb mean free paths \citep[e.g.,][]{Schekochihin_2006}. This feature results in a complex interplay between a plasma's macrophysical evolution (e.g., due to expansion, compression, or large-scale shear) and its microphysical response (e.g., departures from local thermodynamic equilibrium, triggering of kinetic instabilities) \citep{Schekochihin_2005,Kunz2014_b,Hellinger_2015a,Riquelme_2015,Sironi_2015,Squire_2017,Kunz_2020}. This interplay becomes increasingly complex when that macrophysical evolution induces or accompanies a cascade of turbulent fluctuations down to microphysical scales, a situation thought to be ubiquitous in the solar wind, low-luminosity black-hole accretion flows, and the intracluster medium \citep[e.g.,][]{Alexandrova_2013,Yuan_2014,Simionescu2019}.

In this paper, we investigate to what extent the basic building blocks of strong, incompressible, Alfv\'{e}nic turbulence---namely, the existence of a conservative cascade from large (injection) to small (dissipative) scales, the locality of interactions between turbulent fluctuations, and a scale-by-scale balance between the characteristic linear oscillation time of the fluctuations and their nonlinear interaction time known as ``critical balance'' \citep{Goldreich_1995,Mallet_2015,Schekochihin_2020}---survive when subject to microphysical constraints dictated by the kinetic evolution of a collisionless plasma. Theoretical work describing magnetized turbulence in weakly collisional or collisionless plasma, but adopting a pressure-isotropic background, suggests that these organizing principles endure, with a local, conservative, Alfv\'enic cascade extending from macroscopic scales down to the ion-Larmor scale \citep{Schekochihin_2009}. However, the assumption of an isotropic background pressure is not always justified; instead, the pressure tensor is more naturally anisotropic with respect to the magnetic field, with the evolution of field-parallel and perpendicular pressures influenced by approximate adiabatic invariance of the charged particles. How this pressure anisotropy alters the properties of Alfv\'enic turbulence has been a question of particular interest in recent years \citep[e.g.,][]{Klein_2015,Kunz_2015,Kunz_2018,Markovskii_2019}.

To address this question, we use results from a hybrid-kinetic simulation in which strong Alfv\'enic turbulence is driven in a collisionless, magnetized plasma undergoing steady expansion transverse to a mean magnetic field. This expansion drives pressure anisotropy in the plasma through approximate adiabatic invariance. We find that, despite the consequent decrease in the characteristic linear frequency and  Alfv\'{e}n ratio of the fluctuations, the Alfv\'{e}nic cascade adapts to maintain critical balance. Eventually the plasma becomes unstable to kinetic firehose instabilities, which grow rapidly on ion-Larmor scales, scatter particles, and thereby impede the further production of pressure anisotropy. Even in this state, critical balance of the Alfv\'{e}nic cascade persists, with the majority of the turbulent motions remaining stable.

\section{Theoretical considerations and method of solution}

\subsection{Why expansion?}

Of the various types of macroscopic evolution that a turbulent, collisionless magnetized plasma can undergo, there are two compelling reasons to consider expansion. 

First, plasma expansion on a timescale $\tau_{\rm exp}$ much larger than the inverse cyclotron frequency $\Omega^{-1}_s$ of each particle species $s$ ($\in\{e,i\}$ for an electron-ion plasma) provides a natural way to drive temperature anisotropy, $\Delta_s \equiv T_{\perp s}/T_{\| s} - 1 \ne 0$, where $T_{\perp s}$ ($T_{\parallel s}$) is the field-perpendicular (-parallel) component of the temperature of species $s$. For example, as plasma expands transversely to a mean magnetic (``guide") field, mass and magnetic-flux conservation dictate that the mean number density $n_s$ of each species $s$ and the guide-field strength $B_{\rm g}$ satisfy $n_s,B_{\rm g}\propto L_\perp^{-2}$, where $L_{\perp}$ is the characteristic transverse size of the plasma (taken to be much larger than the thermal Larmor radius $\rho_s$ of each species; the characteristic parallel size $L_\parallel$ is held fixed). Combined with conservation of the first and second adiabatic invariants, {\em viz.}~$T_{\perp s} \propto B_{\rm g}$ and $T_{\| s} \propto (n_s/B_{\rm g})^2$ (\citealt{Chew_1956}; hereafter, CGL), these scalings imply a decreasing $T_{\perp s}$ while $T_{\| s}$ remains approximately constant. Thus, if $\Delta_s = 0$ initially, then it becomes increasingly negative. Simultaneously, the parallel plasma beta parameters, $\beta_{\| s} \equiv 8 \pi n_s T_{\| s}/B_{\rm g}^2$, increase. That the combination $\beta_{\| s}\Delta_s$ grows increasingly negative has two important consequences. First, the effective Alfv\'en speed
\begin{equation}\label{eqn:vAeff}
    v_{\rm A,eff} \equiv v_{\rm A} \biggl(1+\sum_s \frac{\beta_{\|s}\Delta_s}{2}\biggr)^{1/2}
\end{equation}
drops below the conventional Alfv\'en speed $v_{\rm A}$, tending towards zero as $\sum_s\beta_{\|s}\Delta_s \rightarrow -2$ (at which point there is no energetic cost to bending the field). Thus, the effective tension in the magnetic-field lines is reduced, with Alfv\'{e}n waves becoming unstable for $\sum_s\beta_{\parallel s}\Delta_s < -2$ (the ``fluid firehose'' threshold; \citealt{Chandrasekhar_1958,Parker_1958}). Concurrently, when $\beta_{\|s}\Delta_s \lesssim -1$ the plasma becomes unstable to various kinetic instabilities. Of particular pertinence to Alfv\'enic turbulence are instabilities on ion-Larmor scales: 
the kinetic parallel and oblique firehoses \citep{Yoon_1993}. 
For plasma with $\beta_{\|i} \approx 2$--$4$ and Maxwellian electrons, the oblique firehose operates when $\Delta_i \lesssim -1.4 \beta_{\|i}^{-1}$ \citep{Hellinger_2000}, while the growth rate of the (threshold-less) parallel firehose is $\gamma_{\rm f} \gtrsim 10^{-3}\Omega_i$ for $\Delta_i \lesssim -1.1 \beta_{\|i}^{-1}$ \citep{Matteini_2006}. Both effects prompt several questions, including whether critical balance persists during the expansion, how the kinetic instabilities interact with the Alfv\'{e}nic turbulence, and whether the turbulent motions themselves become unstable and disrupt the cascade.

The second reason to consider the problem of expanding Alfv\'enic turbulence is its relevance to the solar wind. A parcel of solar-wind plasma initially located at a large distance $R \gg L_{\perp}, L_{\|}$ from the Sun and moving radially outwards at speed $v_{\rm sw}$ will undergo (approximately linear) expansion on a characteristic timescale $\tau_{\rm exp} = R/v_{\rm sw}$ \citep[e.g.,][]{Matteini_2012}. Expansion is thought to play an important role in various key physical processes in the solar wind, including plasma heating, the generation of turbulence, and kinetic physics such as the production of temperature anisotropy \citep{Velli_1989,Verdini_2007,Chandran_2009,Matteini_2013,Chandran_2019}. There have therefore been many complementary investigations of expanding plasmas in the solar-wind context \citep[e.g.,][]{Grappin_1993,Liewer_2001,Matteini_2006,Camporeale_2010,Hellinger_2015b,Hellinger_2017b,Hellinger_2019,Squire_2020}.

\subsection{Hybrid-kinetic description of expanding Alfv\'{e}nic turbulence} \label{sec:sims}

We adopt a hybrid-kinetic approach to solve for the multi-scale dynamics of Alfv\'{e}nic turbulence in a collisionless, expanding plasma. A non-relativistic, quasi-neutral ($n\equiv n_i=n_e$) plasma with kinetic ions (mass $m_i$, charge $e$) and massless, fluid electrons is threaded by a uniform magnetic field $\bb{B}_{\rm g} = B_{\rm g}\ez$ and subjected to a random, time-correlated, solenoidal driving force $\bb{F}(t,\bb{r})\perp\bb{B}_{\rm g}$. This driving is the same as described in \citet{Arzamasskiy_2019}; it is designed to mimic the action of random inertial forces arising from an anisotropic cascade of turbulent fluctuations at scales larger than the simulation domain. The electrons are assumed to be pressure-isotropic and isothermal with temperature $T_e = T_{i0}$, the initial ion temperature. A fourth-order hyper-resistivity is used to remove magnetic energy at the smallest scales.

The subsequent evolution of this plasma is solved using the second-order--accurate, particle-in-cell code \pegpp (Arzamasskiy et al., in prep.), which is an optimized implementation of the algorithms detailed in \citet{Kunz_2014a}. Well-resolved 3D hybrid-kinetic simulations of Alfv\'enic turbulence are essential for modelling this problem, in particular for simultaneously capturing both the turbulent cascade above and below ion-Larmor scales and the physics of ion-firehose instabilities. That being said, our treatment of the electrons as an isothermal, isotropic fluid precludes any kinetic instabilities driven by electron temperature anisotropy (e.g., the electron firehose; \citealt{Li_2000}). While the properties of inertial-range Alfv\'{e}nic fluctuations and ion-scale firehose instabilities are not expected to be affected appreciably by electron kinetics, it remains an open question as to how electron anisotropy affects the sub-ion-Larmor cascade of kinetic Alfv\'{e}n waves (KAWs; see \S\S 3.6.2, 4.4, 4.5 of \citealt{Kunz_2018}). For now, we simply note that, in the near-Earth solar wind, the electrons' collisional age seems to control the electron temperature anisotropy \citep{Salem_2003} and the total temperature anisotropy at $\beta\gtrsim{1}$ is dominated by protons \citep{Chen_2016}. By modeling only a single ion species (protons), our simulations also preclude some other effects thought to be relevant in the solar wind, e.g., instabilities driven by drifting helium ions~\citep{Verscharen_2019}.

To model the expansion, \pegpp enacts a coordinate transform from a co-moving, non-expanding frame (position vector $\bb{r}$) to the co-moving expanding frame (position vector $\bb{r}'$) using the time-dependent (diagonal) Jacobian transformation matrix $\msb{\Lambda}(t) \equiv \partial \bb{r}/\partial \bb{r}'$, as in the Hybrid Expanding Box (HEB) model of \citet[][appendix A]{Hellinger_2005}. \pegpp solves the following modified versions of Faraday's and Ohm's laws in the expanding frame for the magnetic field $\bb{B}' \equiv \lambda \msb{\Lambda}^{-1}\bb{B}$ and the electric field $\bb{E}'\equiv \msb{\Lambda}\bb{E}$:
\begin{align}
    \pD{t'}{\bb{B}'} &= -c\grad'\btimes\bb{E}' , \\*
    \bb{E}' &= -\frac{\bb{u}'}{c}\btimes\bb{B}' - \frac{T_e}{en'}\grad' n' + (\grad'\btimes\bb{B}')\btimes\frac{\msb{\Lambda}^2\bb{B}'}{4\pi e n'\lambda} , 
\end{align}
where the primed-frame number density $n'\equiv \lambda n$ and ion-flow velocity $\bb{u}'\equiv \msb{\Lambda}^{-1} \bb{u}$, $\lambda \equiv {\rm det}\,\msb{\Lambda}$, and $t'=t$. These fields are used to update the simulation ion-particle positions $\bb{r}'_p=\msb{\Lambda}^{-1}\bb{r}_p$ and velocities $\bb{v}'_p = \msb{\Lambda}^{-1} \bb{v}_p$ via
\begin{align}
    \D{t'}{\bb{r}'_p} &= \bb{v}'_p , \label{eqn:drdt} \\*
    \D{t'}{\bb{v}'_p} &=  \frac{e}{m_i} \, \msb{\Lambda}^{-2} \left[ \bb{E}'(t',\bb{r}'_p) + \frac{\bb{v}'_p}{c}\btimes\bb{B}'(t',\bb{r}'_p)\right] \nonumber\\*
    \mbox{} &+\msb{\Lambda}^{-1}\frac{\bb{F}(t',\bb{r}'_p)}{m_i} - 2 \msb{\Lambda}^{-1}\D{t'}{\msb{\Lambda}}\,\bb{v}'_p . \label{eqn:dvdt}
\end{align}
The final (velocity-dependent) term in equation \eqref{eqn:dvdt} is straightforwardly incorporated into the semi-implicit Boris algorithm for solving particle trajectories alongside the $\bb{v}'_p\btimes\bb{B}'$ rotation. Quantities in the non-expanding frame are easily obtained {\it ex post facto}.

The expansion is taken to be perpendicular to $\ez$ and linear in time: $\msb{\Lambda}(t) = \ez\ez + \left(1+t/\tau_{\rm exp}\right) \bigl(\msb{I}-\ez\ez\bigr) $, where $\tau_{\rm exp}$ is the expansion time and $\msb{I}$ the unit dyadic. Thus the perpendicular size of the simulated plasma increases in time as $L_\perp(t) = L_{\perp 0}(1+t/\tau_{\rm exp})$, while the parallel size of the simulated plasma remains constant, $L_{\|}(t) = L_{\|0}$. (We denote any given quantity $X$ evaluated at the start of the simulation by $X_0$.) Magnetic-flux conservation then gives $B_{\rm g}(t) = B_{\rm g0}(1+t/\tau_{\rm exp})^{-2}$. This prescription is physically relevant to the expanding solar wind at ${\gtrsim}0.1~{\rm au}$, on account of the solar wind's constant speed and radial direction at those distances \citep{Verscharen_2019}, although our treatment of the mean magnetic field as radial is a simplifying assumption.

%
%
\begin{figure}
\includegraphics[width=\linewidth]{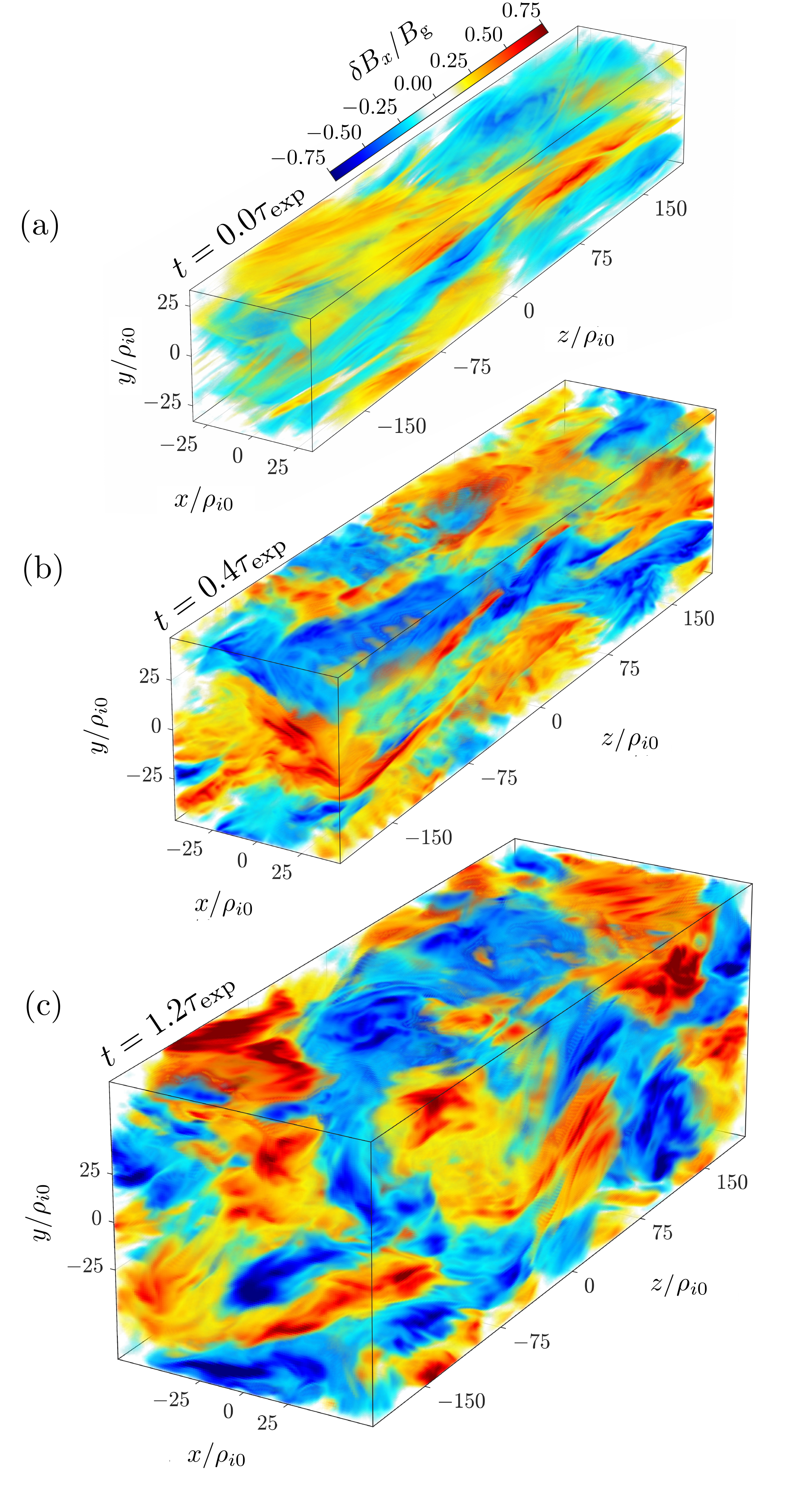}
\centering
\caption{Volume rendering of the $x$ component of the magnetic field, $\delta B_x/B_{\rm g}$, (a) just prior to expansion, (b) when the firehose modes emerge, and (c) near the end of the run well after one expansion time. Regions where $|\delta B_x|/B_{\rm g}$ is small are transparent.
\label{fig:fig1}}
\end{figure}

%
%
\begin{figure*}
\centering
\includegraphics[width=\linewidth]{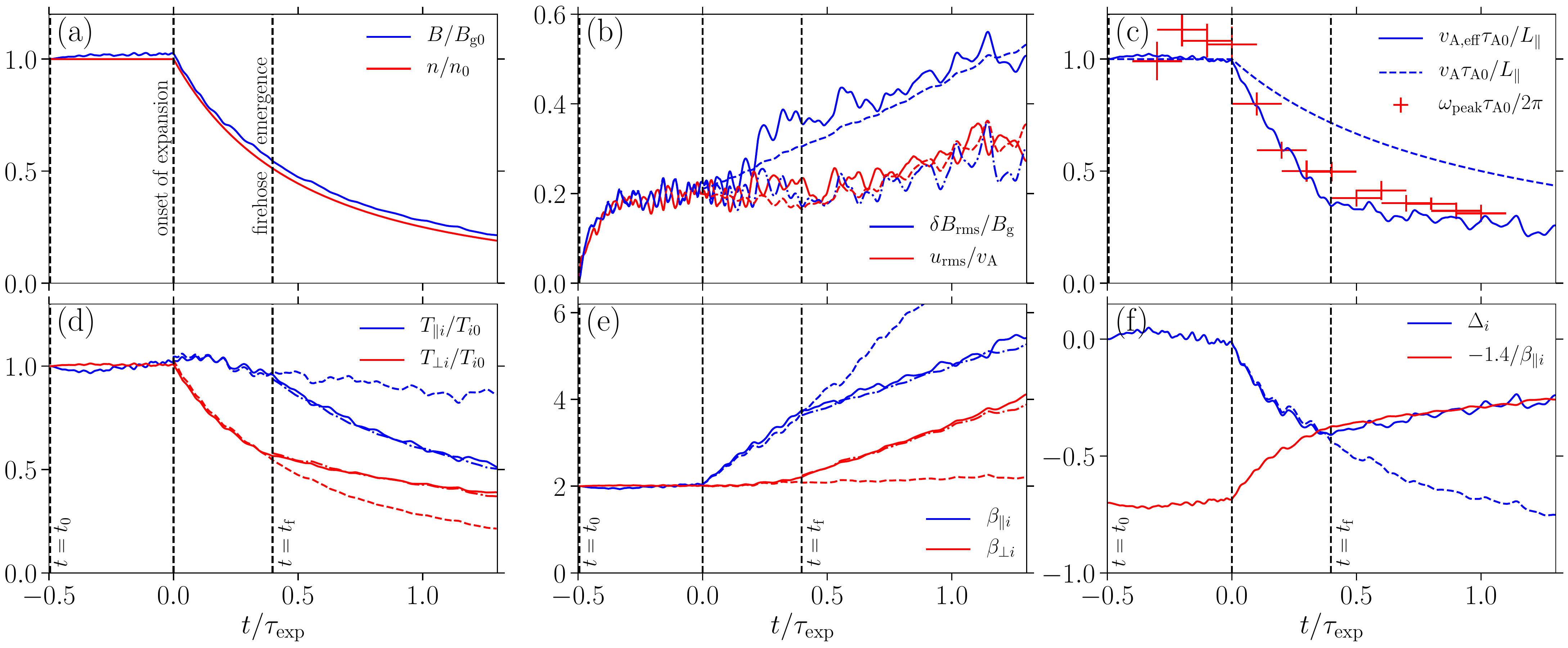}
\caption{(a) Evolution of box-averaged $B$ and $n$, normalized by their initial values. (b) Evolution of $\delta B_{\rm rms}/B_{\rm g}$ and $u_{\rm rms}/v_{\rm A}$ (solid lines), compared with their theoretical expectations (dashed lines; see text). The blue dot-dashed line traces $(1+{\beta}_{\|i}{\Delta}_i/2)^{1/2} \delta B_{\rm rms}/B_{\rm g}$, for which the kinetic-normalized Alfv\'{e}n ratio $r_{\rm A,eff}=1$. (c) Evolution of the spectral-peak frequency $\omega_{\rm peak}$ (normalized by $2\pi/\tau_{\rm A0}$) of the magnetic fluctuations (red pluses), compared to the outer-scale Alfv\'{e}n frequency (blue dashed line) and effective Alfv\'en frequency (blue solid line). Vertical error bars on $\omega_{\rm peak}$ represent standard errors; horizontal error bars represent the size of the Gaussian window function used to obtain the time-dependent frequency spectra. (d) Evolution of box-averaged ${T}_{\perp i}$ and ${T}_{\|i}$, normalized by their initial values (solid lines), with their double-adiabatic predictions (dashed lines) and those from our anomalous collisionality model (dot-dashed lines). (e) Evolution of ${\beta}_{\| i}$ and ${\beta}_{\perp i} \equiv 8 \pi n T_{\perp i}/B_{\rm g}^2$, with their double-adiabatic counterparts (dashed lines) and those from our anomalous collisionality model (dot-dashed lines). (f) Evolution of ${\Delta}_{i}$ (blue solid line) compared with its double-adiabatic prediction (blue dashed line). The (approximate) threshold for the kinetic firehose instability in a bi-Maxwellian plasma, ${\Delta}_{i} = -1.4/{\beta}_{\| i}$ (solid red line), is shown.
\label{fig:fig2}}
\end{figure*}

\subsection{Physical set-up} \label{sec:phys_setup}

At the start of the simulation (time $t_0$), $N_{\rm ppc}=10^3$ simulation ion-particles per cell are drawn randomly from a stationary Maxwellian distribution with temperature $T_{i0}$ and number density $n_0$ and placed uniformly in an elongated 3D computational domain of size $L_x \times L_y \times L_z = (65\rho_{i0})^2\times 390\rho_{i0}$ containing $256^2\times 1536$ cells. At this box size and resolution, the captured wavenumbers are initially in the range $k_{(x,y)}\rho_{i0}\in[0.097,12.37]$ and $k_z\rho_{i0}\in[0.016,12.37]$. The initial ion beta parameter is $\beta_{\|i0} = 2$, representative of near-Earth conditions in the solar wind \citep{Matteini_2007}. Prior to initiating expansion, steady-state Alfv\'enic turbulence is generated in the plasma by forcing the particles with an $\bb{F}(t,\bb{r})$ having the correlation time $\tau_{\rm A0}/2\pi$, where $\tau_{\rm A0} \equiv L_z/v_{\rm A0}\approx 552\Omega^{-1}_{i0}$ is the initial Alfv\'{e}n-crossing time, $v_{\rm A0} \equiv B_{\rm g0}/(4\pi m_i n_0)^{1/2}$ is the initial Alfv\'{e}n speed, and $\Omega_{i0} \equiv eB_{\rm g0}/m_i c$ is the initial ion-cyclotron frequency. The magnitude of the force is such that critical balance is maintained for the box-scale fluctuations: $u_{\rm rms}/v_{\rm A0} \approx L_\perp/L_\|$, where $u_{\rm rms}$ is the root-mean-square (rms) turbulent velocity. Assuming a $-5/3$ power-law scaling for turbulent fluctuations on scales larger than the box, the inferred perpendicular wavenumber at which the energy of the turbulent fluctuations becomes comparable to that of the guide magnetic field is $k_{\perp}^{\rm outer} \sim 10^{-3} \rho_{i0}^{-1}$, a comparable degree of separation to that observed in the fast, $\beta\gtrsim 1$ solar wind \citep{Wicks_2010}.
This initial non-expanding phase of the simulation lasts for five Alfv\'{e}n-crossing times until $t = 0$, so that $t_0 = -5\tau_{\rm A0}\approx -2758\Omega^{-1}_{i0}$. The turbulent magnetic fields at $t=0$ are visualized in Figure~\ref{fig:fig1}(a).

The plasma's expansion is then initiated as described in \S\ref{sec:sims}, with $\tau_{\rm exp} = 10 \tau_{\rm A0} \approx 5515 \Omega_{i0}^{-1}$. This expansion time is comparable to the inferred Alfv\'{e}n-crossing time at the outer scale of the turbulence, similar to conditions in the fast solar wind \citep{Wicks_2010,Alexandrova_2013}. It also means that the turbulent heating time $\tau_{\rm heat} \sim (3/2) T_i L_{\perp}/(m_i u_{\rm rms}^3) \gtrsim 5 \tau_{\rm exp}$ in our simulation; as a result, the thermodynamic evolution of the plasma is dominated not by turbulent heating but rather by the approximately double-adiabatic expansion and the feedback from firehose instabilities. 

As the plasma expands, strong Alfv\'{e}nic turbulence is driven continuously such that $u_{\rm rms}(t)\approx L_\perp(t) v_{\rm A}(t)/L_\parallel=\const$. In retrospect, our results suggest that a more realistic forcing prescription would maintain critical balance adaptively at the outer scale using $v_{\rm A,eff}$ instead of $v_{\rm A}$. However, this prescription requires \textit{a priori} knowledge of the temperature anisotropy's evolution to evolve $v_{\rm A,eff}(t)$ and, moreover, becomes problematic if $v_{\rm A,eff}$ were to approach $0$. In practice, we find the only consequence of using $v_{\rm A}(t)$ to determine the forcing amplitude to be a slight excess of energy in the turbulent fluctuations at the outer scale.

\section{Results} \label{sec:results}

The overall plasma evolution is summarized in Figure~\ref{fig:fig2}. Panel (a) shows that the box-averaged magnetic-field strength ${B}(t)$ and density $n(t)$ decrease in tandem once the expansion begins, with ${B}(t)/B_{\rm g0} \approx {n}(t)/n_0 = (1+t/\tau_{\rm exp})^{-2}$. The rms field strength actually decreases slightly slower due to the growth of the turbulent Alfv\'enic fluctuations, $\delta B_{\rm rms}$, relative to $B_{\rm g}(t)$ as the plasma expands (Figure~\ref{fig:fig2}(b)). This growth, also evident in Figure~\ref{fig:fig1}, is caused by wave-action conservation and by the build-up of residual magnetic energy in the fluctuations from the reduced energetic cost of bending field lines in a plasma with $\Delta_i < 0$. Namely, the Alfv\'en ratio $r_{\rm A} \equiv 4 \pi m_i n u_{\rm rms}^2/\delta B_{\rm rms}^2$ becomes smaller than unity as the expansion proceeds, an effect that may be compensated by instead using the ``kinetic normalization'' $r_{\rm A,eff} \equiv r_{\rm A}(1+\beta_{\| i}\Delta_i /2)^{-1}$ \citep{Chen_2013}. The associated relation $\delta B_{\rm rms}/B_{\rm g} \approx (1+\beta_{\| i}\Delta_i/2)^{-1/2}\, u_{\rm rms}/v_{\rm A}$, when combined with critical balance of the box-scale fluctuations, {\em viz}.~$u_{\rm rms} \sim [L_{\perp}(t)/L_{\|}] v_{\rm A, {eff}}(t) \propto (1+t/\tau_{\rm exp})[1+\beta_{\|i}(t)\Delta_i(t) /2]^{1/2}$ (Figure~\ref{fig:fig2}(b), red-dashed line), implies $\delta B_{\rm rms}/B_{\rm g} \propto (1+t/\tau_{\rm exp})$ (Figure~\ref{fig:fig2}(b), blue-dashed line), a manifestly good fit to the data.

Another key property of Alfv\'enic turbulence in an expanding collisionless plasma is the decreasing characteristic frequency of the fluctuations. This feature is demonstrated by Figure~\ref{fig:fig2}(c), in which the red pluses track the time evolution of the energetically dominant (``peak'') oscillation frequency of the fluctuations, $\omega_{\rm peak}$.\footnote{$\omega_{\rm peak}$ is computed using time series of high-cadence magnetic-field data recorded during the simulation at 27 fixed points in space. These series are Fourier transformed and the frequencies corresponding to the peaks of their corresponding energy spectra are algebraically averaged. To isolate the peak frequency at a particular time, a Gaussian window function (full-width-half-maximum $\Delta t = 0.2 \tau_{\rm exp}$) centered at that time is applied to each series before Fourier transforming.\label{ftn:wpeak}} While some decrease in $\omega_{\rm peak}$ is caused by the decreasing Alfv\'{e}n speed, $v_{\rm A} = v_{\rm A0} (1+t/\tau_{\rm exp})^{-1}$ (blue-dashed line), it is mostly due to the reduction in the {\em effective} Alfv\'{e}n speed caused by $\beta_{\|i}\Delta_i$ becoming increasingly negative. Indeed, the effective Alfv\'{e}n frequency of the box-scale fluctuations, $2\pi v_{\rm A, eff}/L_{\|}$ (solid-blue line), matches the data well.

The production of negative temperature anisotropy during the expansion is shown in Figure~\ref{fig:fig2}(d). During the initial phase, the parallel (blue line) and perpendicular (red line) ion temperatures evolve approximately double-adiabatically: ${T}_{\perp i}(t) \approx {T}_{\perp i}(0) [{B}(t)/{B}(0)]$ (red-dashed line) and  ${T}_{\| i}(t) \approx {T}_{\| i}(0) [{n}(t)/{n}(0)]^{2} [{B}(t)/{B}(0)]^{-2}$ (blue-dashed line). However, at $t \approx t_{\rm f} \equiv 0.4 \tau_{\rm exp}$, an abrupt change in the evolution of ${T}_{\perp i}(t)$ and ${T}_{\| i}(t)$ occurs, and the double-adiabatic predictions no longer hold. This change is coincident with ${\Delta}_i$ decreasing sufficiently (and ${\beta}_{\|i}$ increasing sufficiently---see Figure~\ref{fig:fig2}(e)) that ${\Delta}_i \lesssim -1.4/{\beta}_{\|i}$ (see Figure~\ref{fig:fig2}(f)), at which point the plasma is unstable to kinetic firehose instabilities. Such firehose fluctuations, visually evident near the ion-Larmor scale in Figure~\ref{fig:fig1}(b), are characterized later in this section.

%
%
\begin{figure}
\centering
\includegraphics[width=\linewidth]{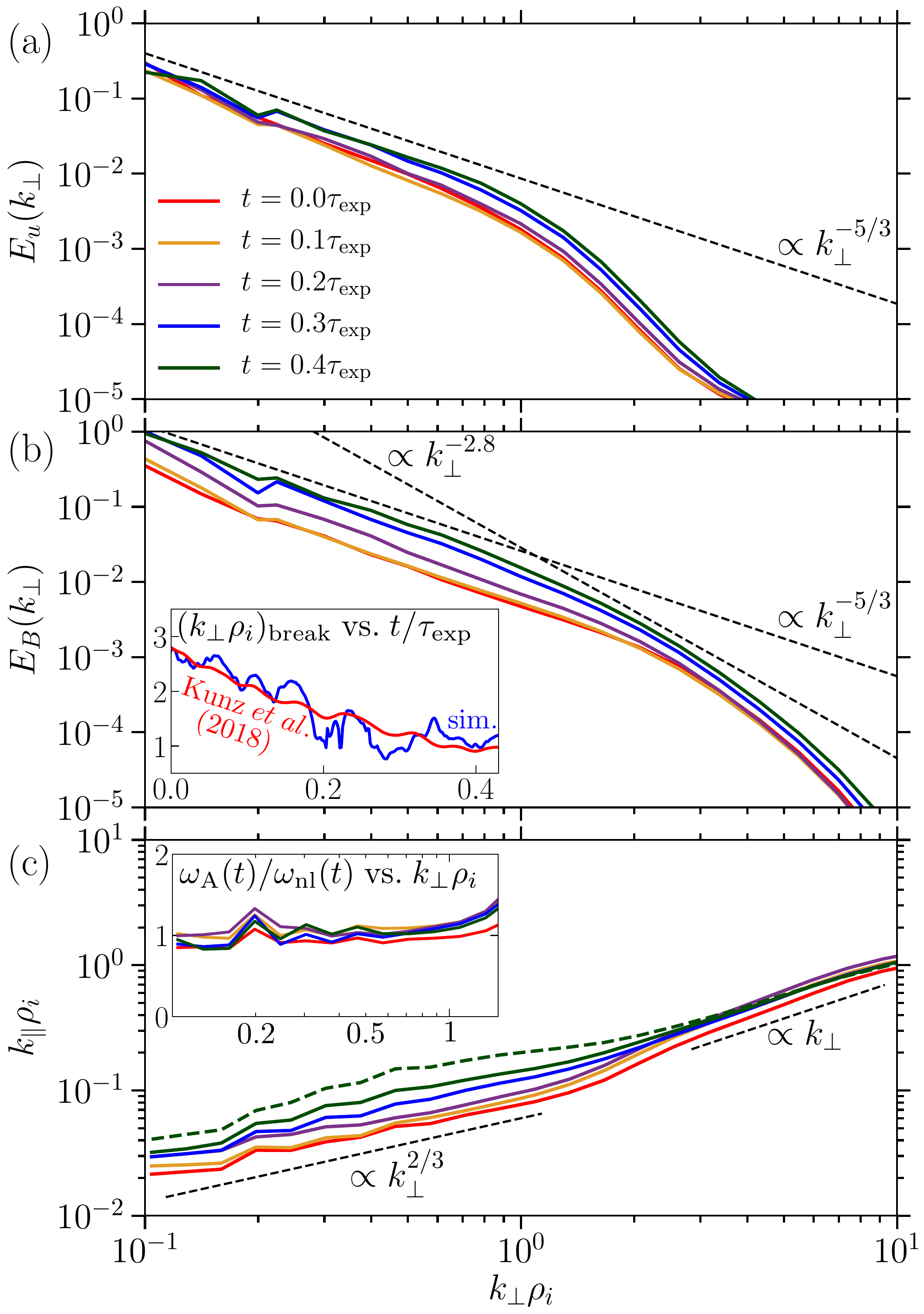}
\caption{Evolution of (a) kinetic and (b) magnetic energy spectra, each obtained by averaging Fourier amplitudes over a time interval of size $\tau_{\rm A0}$. The inset of panel (b) shows the evolution of the spectral break point in the magnetic energy. (c) Instantaneous spatial anisotropy of turbulent fluctuations as a function of perpendicular scale. At $t = 0.4 \tau_{\rm exp}$, the calculation of the anisotropy is weighted towards firehose-stable regions with $\beta_{\|i}\Delta_i\ge -1.4$ (see text); the anisotropy of the full field is denoted by the dashed line. The inset of panel (c) shows the instantaneous ratio of linear Alfv\'en frequency $\omega_{\rm A}(t)$ and nonlinear frequency $\omega_{\rm nl}$ as a function of perpendicular scale. Adaptive critical balance ($\omega_{\rm A}/\omega_{\rm nl} \sim 1$) holds throughout the inertial range.
\label{fig:fig3}}
\end{figure}

Figures \ref{fig:fig3}(a) and (b) display 1D power spectra of the velocity ($E_u$) and  magnetic ($E_B$) fluctuations at select times as functions of the perpendicular wavenumber $k_{\perp}$ normalized to the time-dependent ion-Larmor scale, $\rho_i \equiv [2 T_{\perp i}(t)/m_i]^{1/2}/\Omega_i(t)$. Their overall shapes are similar to those found in prior hybrid-kinetic simulations of non-expanding, $\beta_{\|i} \sim 1$ turbulence \citep[e.g.,][]{Arzamasskiy_2019}: $E_u(k_{\perp}), E_B(k_\perp) \propto k_{\perp}^{-5/3}$ in the inertial (``MHD'') range, before steepening at $k_{\perp} \rho_i \gtrsim 1$ due to finite-Larmor-radius effects. The ``break point'' at which this steepening occurs, $(k_\perp \rho_i)_{\rm break}$ (blue curve, Figure~\ref{fig:fig3}(b) inset), decreases at a rate quantitatively consistent with theoretical expectations \citep[][\S 3.6.4]{Kunz_2018} that
\begin{equation}
    (k_{\perp} \rho_i)_{\rm break} \propto (1+\beta_{\|i}\Delta_i/2)^{1/4} \,\beta_{\|i}^{-1/4}(T_{\perp i}/T_{\|i})^{1/2}
\end{equation}
(red curve, Figure~\ref{fig:fig3}(b) inset).\footnote{The break point $(k_\perp\rho_i)_{\rm break}$ is computed at a given time by first evaluating $\widetilde{E}_{B0} \equiv \int_{k_{\perp \rm l}}^{k_{\perp \rm u}} \rmd k_{\perp} \, k_{\perp}^{5/3} E_B(k_{\perp})/(k_{\perp \rm u}-k_{\perp \rm l})$, where $k_{\perp \rm l}$ and $k_{\perp \rm u}$ define the lower and upper bounds of the inertial range, and then determining the value of $k_{\perp}$ at which $k_{\perp}^{5/3} E_B(k_{\perp})$ falls below some fraction of $\widetilde{E}_{B0}$, denoted by $\widetilde{E}_{B,\mathrm{cut}}$. We use $k_{\perp \rm l} \rho_i = 0.4$, $k_{\perp \rm u} \rho_i = 0.8$, and $\widetilde{E}_{B,\mathrm{cut}} = 0.8 \widetilde{E}_{B0}$; the result is qualitatively insensitive to moderate variations in these parameters.} 
These spectral features are maintained throughout the expansion, even for $t \gtrsim t_{\rm f}$.

Having provided evidence that various properties of the large-scale fluctuations adapt to the changing background pressure anisotropy in a manner consistent with critical balance, we now utilize the spectra in Figure \ref{fig:fig3} to show that critical balance is in fact maintained adaptively, scale by scale, as the plasma expands. We do so by computing the spectral anisotropy of the fluctuations using an approach proposed by \citet{Cho_2009} in which the characteristic parallel wavenumber $k_{\|}(k_{\perp})$ of magnetic-field fluctuations with perpendicular wavenumber $k_{\perp}$ is determined from their rms parallel lengthscale (see their equation (34)). For fluctuations with a given $k_{\perp}$, this measure is most sensitive to the energetically dominant fluctuations with the largest $k_{\|}$, and so the approach can be used to determine the linear frequency $\omega_{\rm A}\equiv k_{\|} v_{\rm A,eff}$ of these fluctuations and compare it with their nonlinear frequency $\omega_{\rm nl} \equiv k_{\perp} [ k_\perp E_u(k_{\perp})+v_{\rm A,eff}^2 k_\perp E_B(k_{\perp})/B_{\rm g}^2 ]^{1/2}$. In critically balanced turbulence, the turbulent energy is concentrated in a cone satisfying $\omega_{\rm A} \lesssim \omega_{\rm nl}$, with the edge of the cone having $k_{\|}\propto k^{2/3}_\perp$ \citep{Goldreich_1995}.

The result of this calculation is shown at different times in Figure~\ref{fig:fig3}(c). At $t=0$, the measured spectral anisotropy in the inertial range is consistent with the critical-balance scaling $k_{\|} \propto k_{\perp}^{2/3}$. As the expansion proceeds, this scaling is maintained as the overall degree of anisotropy decreases in tandem with the decreasing aspect ratio of the plasma. Furthermore, the inset shows that $\omega_{\rm A} \approx \omega_{\rm nl}$ scale by scale; thus critical balance holds adaptively. At $t \approx t_{\rm f}$, firehose modes (which, unlike the Alfv\'enic fluctuations, are not highly elongated in the field-parallel direction) emerge and bias slightly the calculated scaling of $k_{\|}(k_\perp)$ in the inertial range. To mitigate this bias, a weight function is applied to the magnetic field that preferentially removes firehose-unstable regions before evaluating $k_{\|}$. Using this weight function, adaptive critical balance of the Alfv\'{e}nic cascade is seen to persist.\footnote{The weight function at a given time $t$ is constructed by first identifying all cells in which, when time averaged over an interval of size $\tau_{\rm A0}/2$ prior to time $t$, the firehose instability parameter $\beta_{\|i}\Delta_i \leq -1.4$. These regions are then masked, with the edges of the mask smoothed by a Gaussian filter of scale $4 \pi \rho_i$.}

In summary, no dramatic alterations to the fundamental nature of the Alfv\'enic turbulence are observed during expansion, even when kinetic-scale firehose modes are present. Importantly, there is no noticeable destabilization of the inertial-range Alfv\'enic cascade. This result is due to the efficient regulation of the box-averaged temperature anisotropy, which (as shown in Figure~\ref{fig:fig2}(f)) barely drops below ${\Delta}_i \approx -1.4/{\beta}_{\|i}$. While this value of ${\Delta}_i$ is negative enough to destabilize the plasma to  kinetic firehose instabilities, it is above the ``fluid'' firehose instability threshold $\Delta_i = -2/\beta_{\|i}$ below which $v^2_{\rm A,eff}\le 0$ and Alfv\'{e}n waves cease to propagate.

%
%
\begin{figure*}
\centering 
\includegraphics[width=\linewidth]{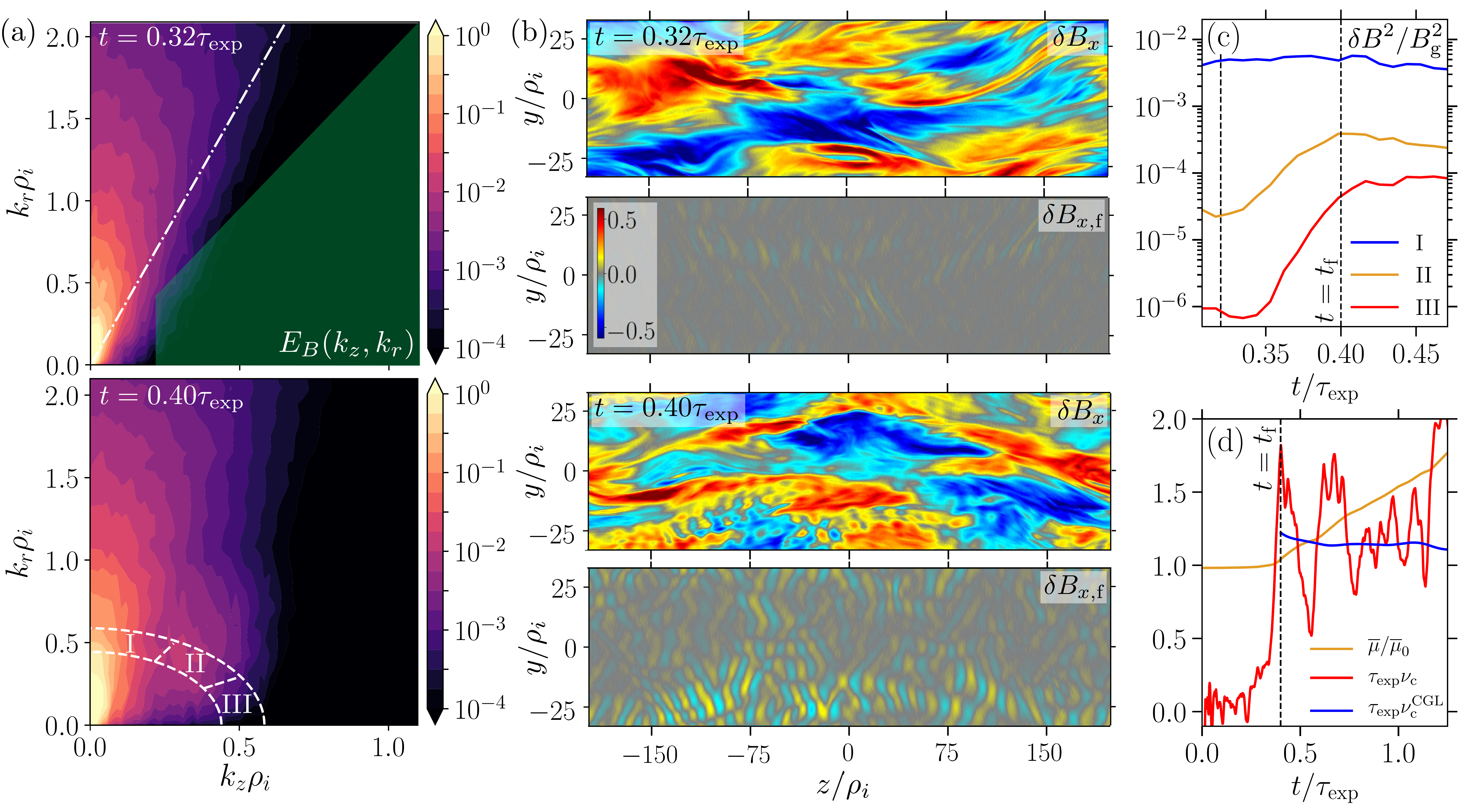}
\caption{(a) Fourier spectrum of magnetic-field fluctuations in $(k_{r},k_{z})$-space at $t = 0.32 \tau_{\rm exp}$ and $0.40 \tau_{\rm exp}$. The Alfv\'{e}nic cascade is spectrally anisotropic, with $k_z\lesssim k_r \tan{\theta_{\rm A}}\approx 0.34 k_r$ (white dot-dashed line); firehose fluctuations emerge in regions II and III. (b) 2D slice of $\delta B_x$ and its ``firehose'' part $\delta B_{x,\mathrm{f}}$ at the same times in the plane $x = L_{\perp}/2$ (cf.~the right-hand face of the box in Figure~\ref{fig:fig1}(b)). The Fourier-space mask used to separate out the firehose part is indicated in panel (a) by the shaded region. (c) Evolution of magnetic energy for fluctuations with $k \in [0.85,1.15] k_{\rm f}$  (where $k_{\rm f}$ is the firehose wavenumber predicted from linear theory) in three different wavevector-angle bins (measured with respect to the guide field and labelled ${\rm I}$, ${\rm II}$, ${\rm III}$ in panel (a), bottom). (d) Evolution of box-averaged first adiabatic invariant (orange line), effective collisionality $\nu_{\rm c}$ (red line), and model collisionality $\nu_{\rm c}^{\rm CGL}$ for $t > t_{\rm f}$ (blue line).
\label{fig:fig4}}
\end{figure*}

The character of the kinetic-scale firehose fluctuations can be ascertained by examining the 2D Fourier spectrum of the magnetic field $E_B$ in $[k_{r}\equiv (k_x^2+k_y^2)^{1/2},k_{z}]$-space. At $t \lesssim 0.32 \tau_{\rm exp}$ (Figure~\ref{fig:fig4}(a), top), spectral power is concentrated in the region of $(k_{r},k_{z})$-space that satisfies $k_{z} \ll k_{r}$, affirming the quasi-perpendicular nature of the Alfv\'enic cascade. By $t = 0.4 \tau_{\rm exp}$  (Figure~\ref{fig:fig4}(a), bottom), an additional region with spectral power is clearly visible, with its centroid located at $(k_{r} \rho_i,k_{z}\rho_i) \approx (0.4,0.3)$. We associate this power with growing oblique firehose fluctuations.\footnote{In principle, parallel firehose fluctuations sitting atop local field-line deformations caused by the Alfv\'{e}nic turbulence could also appear as oblique modes in $(k_{r},k_{z})$-space. However, the characteristic angular deviation of the magnetic-field lines associated with the Alfv\'enic turbulence is relatively small ($\theta_{\rm A}\approx 19^\circ$), while the observed modes have $\theta\approx 53^\circ$. We thus conclude that the emergent region of spectral power seen in Figure~\ref{fig:fig4}(a) at $t=0.40\tau_{\rm exp}$ is caused by the oblique firehose instability.} These fluctuations can be visualized by isolating the ``firehose'' part $\delta B_{x,{\rm f}}$ of the magnetic field using a Fourier-space mask that filters out quasi-perpendicular modes; the region of $(k_{r},k_{z})$-space identified as the firehose part is indicated by the shaded region in Figure~\ref{fig:fig4}(a). While the Alfv\'enic turbulence does not evolve qualitatively during the time interval $t/\tau_{\rm exp} \in [0.32,0.40]$, the firehose fluctuations increase their amplitudes significantly (see Figure~\ref{fig:fig4}(b)). Figure~\ref{fig:fig4}(c), which shows the evolution of the magnetic energy of ion-Larmor-scale modes at different angles to the guide field, confirms that oblique firehose modes are unstable, with maximum growth rate comparable to that predicted by linear theory at $\beta_{\perp i}\approx 3.6$ and $\Delta_i \approx -0.4$, viz.~$\gamma_{\mathrm{f}\perp} \approx 0.02 \Omega_i \approx 120 \tau_{\rm exp}^{-1}$ at $(k_{\rm f\perp}\rho_i,k_{\rm f \|}\rho_i) \approx (0.4,0.3)$. Parallel firehose modes (measured in region III of Figure~\ref{fig:fig4}(a)) are also unstable, but they have a significantly smaller amplitude than the oblique modes.

The firehose fluctuations efficiently regulate the temperature anisotropy, even though their saturated rms magnetic-field strength is much smaller than that of the Alfv\'enic fluctuations at equivalent wavenumbers. They do so by pitch-angle scattering the ions so that the particles' first adiabatic invariants ($\mu \equiv m_i v_{\perp}^2/2B$, where $v_{\perp}$ is the peculiar perpendicular velocity) are no longer conserved (see Figure~\ref{fig:fig4}(d), orange line). The effective collisionality of this anomalous scattering, $\nu_{\rm c}$, may be estimated using the relation $\dot{\overline{\mu}} = - \nu_{\rm c} (\overline{\Delta T_i/B})$, where the overline denotes a box average, $\Delta T_i \equiv T_{\perp i}-T_{\| i}$, and $\dot{\overline{\mu}}$ is the rate of change of $\overline{\mu} \equiv \overline{T_{\perp i}/B}$. Figure~\ref{fig:fig4}(d) indicates that $\tau_{\rm exp}\nu_{\rm c} \ll 1$ for $t < t_{\rm f}$ (i.e., $\mu$ is approximately conserved pre-firehose), while $\tau_{\rm exp}\nu_{\rm c} \sim 1$ for $t \gtrsim t_{\rm f}$ (i.e., $\mu$ is significantly broken by the firehose fluctuations).

A simple model for $\nu_{\rm c}$ may be constructed by adopting three assumptions: (i) that $n(t)/n_0 \approx {B}(t)/{B}_{\rm g 0}$; (ii) that contributions from heat fluxes and turbulent heating to the temperature anisotropy are negligible (the latter being because $\tau_{\rm heat} \gg \tau_{\rm exp}$; see \S\ref{sec:phys_setup}); and (iii) that $\beta_{\|i}\Delta_i \approx \const$ after $t=t_{\rm f}$. Under these conditions, the CGL equations (including collisions) become $\rmd\ln(T_{\perp i}/B)/\rmd t =-\nu_{\rm c}(T_{\parallel i}/T_{\perp i})\Delta_i$ and $\rmd\ln(T_{\parallel i}B^2/n^2)/\rmd t \approx \rmd\ln T_{\parallel i}/\rmd t = 2\nu_{\rm c}\Delta_i$. The third assumption then implies $\nu_{\rm c} \approx  (3 \Delta_i)^{-1}\,\rmd\ln B/\rmd t \equiv \nu^{\rm CGL}_{\rm c}$. The agreement between this model (Figure~\ref{fig:fig4}(d), blue line) and $\nu_{\rm c}$ evaluated directly from the simulation is good, although $\nu_{\rm c}$ fluctuates significantly. A direct calculation of the mean $\mu$-breaking time of ${\sim}10^4$ tracked particles, following \citet{Kunz2014_b,Kunz_2020} and \citet{Squire_2017}, yields an effective collisionality ${\simeq}\nu^{\rm CGL}_{\rm c}$ for $t \gtrsim t_{\rm f}$. Setting $\nu_{\rm c}=\nu^{\rm CGL}_{\rm c}$ in the above equations leads to a simple equation for the parallel temperature, $\rmd\ln{(T_{\| i}/B^{2/3})}/\rmd t = 0$, so that $T_{\| i}(t) \approx T_{\|i}(t_{\rm f}) [B(t)/B(t_{\rm f})]^{2/3}$. Further setting $\Delta_i \approx -1.4/\beta_{\|i}$ yields $T_{\perp i}(t) \approx T_{\|i}(t)-1.4 B_{\rm g}^2(t)/8 \pi n(t)$. This model is plotted in Figure~\ref{fig:fig2}(d); given its simplicity, its agreement with the actual result is remarkable.

%
%
\begin{figure*}
\centering
\includegraphics[width=\linewidth]{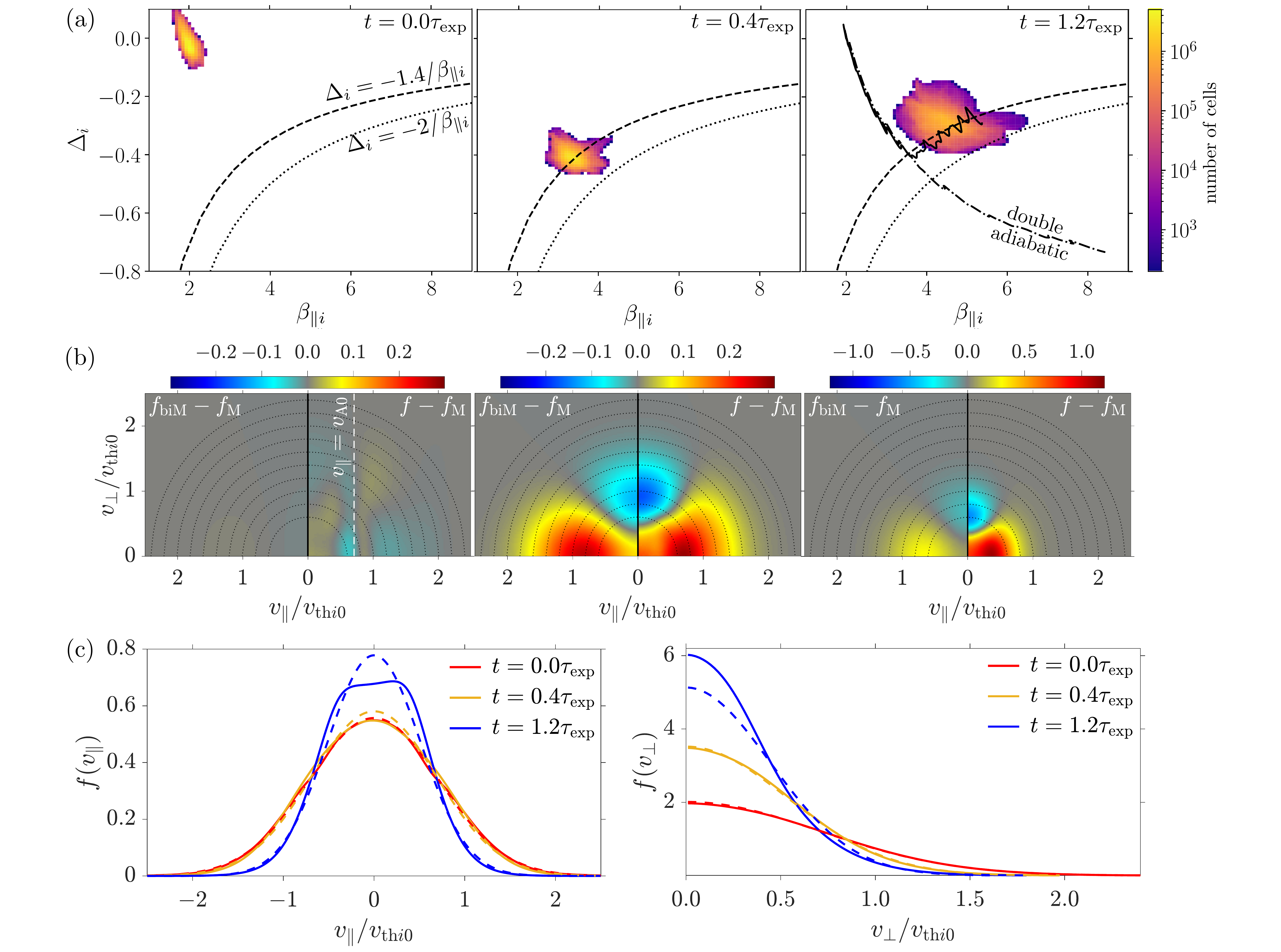}
\caption{(a) PDF of data in $(\beta_{\|i},\Delta_i)$ phase-space at $t = 0$ (left), $t = t_{\rm f}$ (middle), and $t = t_{\rm f} + 0.8 \tau_{\rm exp}$ (right). For each panel, $\beta_{\|i}$ and $\Delta_i$ are averaged over a time interval of $100 \Omega_{i0}^{-1} \sim \gamma_{\rm f \perp}^{-1}$ and spatially averaged (using a Gaussian filter) over a scale $4 \pi \rho_i \sim 2 \pi /k_{\rm f \perp}$. The phase-space trajectory of $(\beta_{\i},\Delta_i)$ associated with Figure \ref{fig:fig2}(e,f) is traced by the solid line; its double-adiabatic counterpart is traced by the dot-dashed line. (b)  $(v_{\|},v_{\perp})$-space plots at the same times of: (right-hand side of each plot) the difference between the (gyro-averaged) ion distribution function $f$ and a Maxwellian distribution function $f_{\rm M}$ with the same temperature; and (left-hand side of each plot) the difference between $f_{\rm M}$ and a bi-Maxwellian distribution function $f_{\rm biM}$ with the same parallel and perpendicular temperatures as $f$. All distribution functions are normalized so that $\int_{-\infty}^{\infty} \rmd v_{\|} \int_{0}^{\infty} \rmd v_{\perp} \,v_{\perp} f = 1$, with $v_{\|}$ and $v_\perp$ being the peculiar parallel and perpendicular velocities. The dashed line on the left panel indicates $v_{\|} = v_{\rm A0}$. 
(c) Parallel ($f(v_{\|})$) and perpendicular ($f(v_{\perp})$) distribution functions at the same times. Dashed lines denote the corresponding $f_{\rm biM}$.
\label{fig:fig5}}
\end{figure*}

The regulation of temperature anisotropy can be elucidated further by considering PDFs of the simulation data in the $(\beta_{\|i},\Delta_i)$ phase space \citep[e.g.,][]{Bale_2009}. Figure~\ref{fig:fig5}(a) shows these PDFs at different stages: at the expansion's start ($t=0$), at $t = t_{\rm f}$, and more than a full expansion time after $t = 0$; the phase-space trajectory of the PDF's average is indicated in the final panel by the black solid line. In all three cases, the relatively small dispersion in $\beta_{\|i}$ and $\Delta_i$ is consistent with the small rms amplitude of the turbulent fluctuations. The temperature anisotropy clearly approaches the oblique firehose instability threshold $\Delta_i = -1.4/\beta_{\|i}$ (dashed line) and subsequently evolves along marginal instability.

Despite the success of our collisionality model, the compartmentalization of all of the kinetic physics into an effective collision frequency hides some interesting emergent features in the ion distribution function $f(v_\parallel,v_\perp)$. Figure~\ref{fig:fig5}(b) shows the difference between $f$ and a Maxwellian distribution with the same temperature as $f$ at three different times during the expansion (with all velocities normalized by the initial thermal speed $v_{{\rm th}i0} \equiv (2 T_{i0}/m_i)^{1/2}$). For comparison, the difference between $f$ and a bi-Maxwellian distribution with the same values of $T_{\| i}$ and $T_{\perp i }$ as $f$ is also shown. Prior to the start of the expansion, the slight deficit of particles with (peculiar) parallel velocities $v_{\|}$ just below the Alfv\'en velocity $v_{\rm A}$ (Figure~\ref{fig:fig5}(b), left panel) is indicative of collisionless damping of the (kinetic) Alfv\'enic fluctuations. Once the expansion begins, these deviations are dwarfed by the expansion-driven temperature anisotropy (Figure~\ref{fig:fig5}(b), middle panel), which, on account of approximate double-adiabaticity, causes $f$ to look like a bi-Maxwellian. However, by late times in the simulation, significant deviations from a bi-Maxwellian are evident (Figure~\ref{fig:fig5}(b), right panel), a finding seen in previous studies of the firehose instability \citep[e.g.,][]{Hellinger_2017b}. In particular, the distribution function integrated over perpendicular velocities, $f(v_\parallel) \equiv \int_{0}^{\infty} \rmd v_{\perp} v_{\perp} f$, exhibits a flattened core (Figure~\ref{fig:fig5}(c)); the distribution function integrated over parallel velocities, $f(v_\perp) \equiv \int_{-\infty}^{\infty} \rmd v_{\|} f$, shows that the anisotropy of the distribution function at subthermal velocities is much more pronounced than in a bi-Maxwellian. These features can be attributed to resonant interactions between ions and the oblique firehose modes (Bott et al., in prep.).

\section{Discussion} \label{sec:discussion}

That the nonlinear interactions between Alfv\'{e}nic fluctuations adapt to satisfy critical balance, even as the characteristic linear frequency of those fluctuations is reduced by pressure anisotropy, is a vivid illustration of the complex interplay between velocity space and configuration space that is central to collisionless plasma physics. This interplay is made richer at $\beta\gtrsim{1}$ by the emergence of ion-Larmor-scale firehose fluctuations, which establish a direct link between the microscales and macroscales by regulating the pressure anisotropy and thereby controlling the effective tension of magnetic-field lines. Despite the small-scale injection of magnetic energy by the firehose, those fluctuations are not sufficient in amplitude to contribute significantly to the magnetic power spectrum (at least perpendicular to the guide field). This finding should ease the concern expressed in \citet{Bale_2009} that ``these local [kinetic] instabilities \dots may confuse the interpretation of solar wind magnetic power spectra''. From the standpoint of the Alfv\'{e}nic cascade, the most important (and potentially observable) roles played by the firehose are as a direct regulator of pressure anisotropy and an indirect mediator of adaptive critical balance and the transition to the KAW range.

The evolution of purely decaying, magnetized turbulence in an expanding, collisionless plasma with $\beta_{\|i} \gtrsim 1$ was recently investigated by \citet{Hellinger_2019} using HEB simulations. In their set-up, an isotropic spectrum of Alfv\'enically polarized waves (amplitude $\delta B_{\rm rms}/B_{\rm g} = 0.24$) was initiated inside a cubic simulation domain with $512^2\times 256$ cells spanning $L^2_\perp\times L_\parallel = (82 \rho_{i0})^3$, before transverse expansion was introduced ($\tau_{\rm exp} = 10^4\Omega_{i0}^{-1}$) and the system evolved. The initial ion distribution function had non-zero temperature anisotropy, $\Delta_{i0} = -0.25$, with $\beta_{\|i0} = 2.4$. Where there is overlap with their results, we find agreement: efficient regulation of the temperature anisotropy by kinetic firehose instabilities, persistence of a quasi-perpendicular Alfv\'enic cascade independent of firehose fluctuations, and distortion of the particle distribution function away from a bi-Maxwellian. There are, however, two important distinctions worth highlighting. First, because of the shape of the simulation domain ($L_{\|} \leq L_{\perp}$) in \citet{Hellinger_2019}, the Alfv\'enic fluctuations are likely not in critical balance. Alfv\'enic fluctuations in an MHD turbulent cascade become critically balanced for isotropic outer-scale fluctuations at a scale $\lambda_{\rm CB} \sim L_{\|} (\delta B_{\rm rms}/B_{\rm g})^{3/2}$ \citep{Schekochihin_2020}; given the parameters in \citet{Hellinger_2019}, we estimate $\lambda_{\rm CB} \approx 0.1 L_{\|} \sim \rho_i$, placing the entire inertial range in the weak-turbulence regime. Our demonstration of adaptive critical balance of strong Alfv\'{e}nic turbulence when the distribution function is anisotropic (even unstably so) is one of our key results. Secondly, we followed the evolution of the turbulence for well over an expansion time, and so could confirm that the temperature anisotropy remains pinned to the kinetic firehose instability threshold as the expansion proceeds. This is an important result for solar-wind applications, because the expansion time there is comparable to the turnover time (and thus the characteristic decay time) of the outer-scale turbulent eddies.

Our conclusions may not hold for plasmas with much higher $\beta_{\|i}$ than have been considered here. First, it is possible to show using linear theory that, if $\tau_{\rm exp}\lesssim 10 \beta^{3/2}_{\parallel i} (\ln\beta_{\parallel i})^{1/2} \Omega^{-1}_i$, then $\Delta_i$ would not be regulated fast enough by the oblique firehose to remain ${>}-2/\beta_{\|i}$. In this case, $v^2_{\rm A,eff}$ would pass through $0$ and the entire inertial-range Alfv\'{e}nic cascade would be destabilized. For the value of $\tau_{\rm exp}$ used in our simulation, we expect this to occur for $\beta_{\|i}\gtrsim 50$. Secondly, negative pressure anisotropy driven by the Alfv\'{e}nic fluctuations themselves can ``interrupt'' the fluctuations if $\delta B_{\rm rms}/B_{\rm g} \gtrsim \beta^{-1/2}_{\| i}$, by nullifying the restoring tension force and exciting a sea of scattering firehose fluctuations \citep{Squire_2017}. An investigation of strong Alfv\'{e}nic turbulence at such high beta is already underway.

%
%
\acknowledgements

AFAB, MWK, and EQ were supported by DOE awards DE-SC0019046 and DE-SC0019047 made through the NSF/DOE Partnership in Basic Plasma Science and Engineering. Support for LA was provided by the Institute for Advanced Study. Support for JS  was provided by Rutherford Discovery Fellowship RDF-U001804 and Marsden Fund grant UOO1727, which are managed through the Royal Society Te Ap\=arangi. High-performance computing resources were provided by: the Texas Advanced Computer Center at The University of Texas at Austin under grant number TG-AST160068; and the PICSciE-OIT TIGRESS High Performance Computing Center and Visualization Laboratory at Princeton University. This work used the Extreme Science and Engineering Discovery Environment (XSEDE), which is supported by NSF grant OCI-1053575. This work benefited from useful conversations with Silvio Sergio Cerri and Alexander Schekochihin, and especially from contributions by Ryan Golant to the implementation of the expanding box in \pegpp while a Princeton University undergraduate in 2019.

%
%
\bibliographystyle{apj}
\bibliography{biblio}

\end{document}